\documentclass[%
 reprint,
 amsmath,amssymb,
 aps,
 pra,
]{revtex4-2}

\usepackage{graphicx}
\usepackage{dcolumn}
\usepackage{bm}
\usepackage{hyperref}
\usepackage[mathlines]{lineno}
\usepackage{xcolor}
\usepackage{booktabs}

\newcommand{\myincludee}[1]{0}

\usepackage{empheq}

\newcommand{\dif}{\mathrm{d}}
\newcommand{\Var}{\operatorname{Var}}

\begin{document}

\preprint{APS/123-QED}

\title{Fundamental entropic processes in the theory of optical thermodynamics}

\author{Nikolaos K. Efremidis}
 \email{nefrem@uoc.gr}
\affiliation{%
 Department of Mathematics and Applied Mathematics, University of Crete, 70013 Heraklion, Crete, Greece
}%
 \affiliation{Institute of Applied and Computational Mathematics, Foundation for Research and Technology-Hellas (FORTH), 70013 Heraklion, Crete, Greece}

\author{Demetrios N. Christodoulides}
\affiliation{CREOL/College of Optics, University of Central Florida, Orlando, Florida 32816, USA}%

\date{\today}

\begin{abstract}
  We study the statistical behavior of multimoded optical systems under equilibrium conditions. We investigate the role of variations of the system parameters in the thermodynamic description and derive, an optical analogue of the first law of thermodynamics, a generic expression for the work done to the system, and an optical Gibbs-Duhem equation. To demonstrate these effects, we focus in the case of two-dimensional photonic lattices. We study the conditions under which the entropy in such waveguide arrays can be considered as extensive. In this respect, small deviations from the extensive character of the entropy give rise to stress and strain terms. We examine how the conservation laws in such array configurations are affected by variations in the system parameters, and furthermore, we analyze the respective thermodynamic processes (isentropic and Joule-type expansions).
\end{abstract}

\maketitle

\section{Introduction}

Optical systems are by nature capable of supporting a multitude of modes with respect to their allowed degrees of freedom. 
In the spatial domain, such structures include multimode fibers~\cite{richa-np2013}, and multicore waveguide arrays~\cite{chris-nature2003}, while in the frequency realm these could be cavity and microcavity arrangements~\cite{vahal-nature2003}.
Over the last few years, multimoded optical systems have attracted considerable attention, both because of their scientific interest and potential applications. In this respect, multimode optical fibers are nowadays intensely investigated, in order to substantially scale-up the information carrying capacity of communication networks through the exploitation of the modal degrees of freedom~\cite{richa-np2013}. In addition, multimoded fiber structures can be utilized in high-power settings like spatio-temporal mode-locking~\cite{wrigh-science2017,wrigh-np2020}, and supercontinuum generation~\cite{wrigh-np2015,lopez-ol2016,krupa-ol2016,longh-ol2003,eftek-nc2019}.
In these latter studies, a rather surprising and unexpected phenomenon was observed in nonlinear multimode graded-index fibers--the so-called ``beam self-cleaning'' effect~\cite{lopez-ol2016,liu-ol2016,krupa-np2017}. During this process, a highly multimode and spatially irregular beam evolves towards a bell-shaped speckle-free pattern, an effect that cannot be explained by invoking well-known nonlinear mechanisms like that of self-focusing, stimulated Raman, and Brillouin scattering.
This issue was addressed in a work published last year, where it was recognized that, although some kind of nonlinearity is necessary for achieving four-wave mode mixing, the actual origin of this effect is thermodynamic~\cite{wu-np2019}: During propagation, with the aid of a moderate nonlinearity, the system reaches thermal equilibrium. 

The theoretical formalism developed in~\cite{wu-np2019} provides a self-consistent thermodynamic approach that can be applied to a variety of multimoded nonlinear optical settings. Specifically, it was shown that such systems can reach thermal equilibrium by maximizing their entropy according to the second law of thermodynamics. 
Once thermalized, the statistics of the modal occupancies obey a Rayleigh-Jeans distribution that is characterized by an optical temperature $T$ and a corresponding chemical potential $\mu$. An optical ``pressure'' $P$ was introduced as the conjugate variable to the number of modes $M$. 
The conditions governing $T$ and $\mu$ were analyzed in~\cite{parto-ol2019}, and in~\cite{makri-ol2020,ramos-prx2020} the thermodynamic laws were derived by employing a grand-canonical approach.  

Previous research activity in this area has been primarily focused on non-equilibrium kinetic formulations based on wave turbulence~\cite{picoz-pr2014,picoz-prl2005,picoz-oe2007,sun-np202,chioc-epl2016,baudi-prl2020}. A number of other publications has also examined phase transitions in discrete lattices based on local mode descriptions~\cite{rasmu-prl2000,silber-prl2009,small-pra2011,kotto-pre2011,derev-pra2013,buons-pra2017,levy-prb2018}.

In this work, we examine how variations in the parameters of a weakly nonlinear multimoded optical system affect its thermodynamic properties. In this respect, we derive an optical analogue of the first law of thermodynamics, which describes all possible sources that cause variations in the internal energy. We find that work can be imparted on the system either by changing the number of modes or by changing the specific parameters of each setting. In addition, we derive a Gibbs-Duhem type equation that relates the variations in the intensive parameters. To demonstrate our predictions, we study the case of two-dimensional photonic lattices arranged in a rectangular geometry. Here, the role of the system parameters is played by the coupling coefficients. Stress and strain terms are introduced to account for corrections due to system asymmetries. The significance of these terms is related to the extensivity of the entropy, which we examine numerically. Finally, we study two particular examples of thermodynamic processes, namely isentropic and Joule-type expansions, and compare the numerical results with theoretical predictions.

\section{Effect of variation of parameters in the thermodynamic description of multimoded optical systems\label{sec:11}}

Our first step is to develop a thermodynamic formalism that takes into account the system parameters of a multimoded optical setting. 
In this respect, let us consider a multimoded system with $M$ modes. In modal space the wavefunction can be expanded as
\[
  |\Psi\rangle =\sum_{l=1}^MC^{(l)}(z)|l\rangle,
\]
so that $\hat H|l\rangle=\varepsilon^{(l)}|l\rangle$, 
where $\hat H$ is the Hamiltonian, $|l\rangle$ is the eigenmode with index $l$ and eigenvalue $\varepsilon^{(l)}$, while $C^{(l)}(z)$ is the respective amplitude that depends on the propagation distance $z$. We consider optical systems that conserve the total power
\begin{equation}
  N = \langle\Psi|\Psi\rangle=\sum_{l=1}^Mn^{(l)}
  \label{eq:N}
\end{equation}
as well as the internal energy
\begin{equation}
  U=\langle\Psi|\hat H|\Psi\rangle   =\sum_{l=1}^M\varepsilon^{(l)}n^{(l)},
  \label{eq:U0}
\end{equation}
where $n^{(l)}=|C^{(l)}|^2$. 
We utilize a grand canonical ensemble that respects these two conservation laws, with $\alpha$, $\beta$ being the corresponding Lagrange multiplies~\cite{pathr-2011}. The resulting probability is given by
\[
p = e^{-q-\alpha \sum_ln^{(l)}-\beta\sum_l\varepsilon^{(l)}n^{(l)}},
\]
where the $q$-potential 
\begin{multline}
  q=
  \sum_{l=1}^M\log
  \int_0^\infty e^{-\alpha n^{(l)}-\beta\varepsilon^{(l)}n^{(l)}}
  \dif n^{(l)}
  = \\
  \sum_{l=1}^M\log\frac1{\alpha+\beta\varepsilon^{(l)}}
  \label{eq:qpot}
\end{multline}
is related to the grand canonical partition function $\mathcal Q$ via $q = \log \mathcal Q$. Note that a grand-canonical formalism was also utilized in~\cite{makri-ol2020,ramos-prx2020}. We assume that the system depends on the additional variables  $\xi=\{\xi_1,\ldots,\xi_J\}$ that are introduced to the partition function through the energy spectrum $\varepsilon^{(l)}=\varepsilon^{(l)}(\xi)$ and thus $q=q(\alpha,\beta,M,\xi)$. 
Taking the differential of Eq.~(\ref{eq:qpot}) we find that the average modal occupation numbers satisfy a Rayleigh-Jeans distribution~\cite{wu-np2019}
\begin{equation}
  \langle n^{(l)}\rangle=
  -\frac1{\beta}
  \left(
    \frac{\partial q}{\partial\varepsilon^{(l)}}
  \right)_{\alpha,\beta,M,\overline{\varepsilon^{(l)}}}=
  \frac{1}{\alpha+\beta\varepsilon^{(l)}},
  \label{eq:RJ}
\end{equation}
as well as
\begin{equation}
  \langle N\rangle
  =
  -
  \left(
    \frac{\partial q}{\partial\alpha}
  \right)_{\beta,\varepsilon,M}
  =
  \sum_{j=1}^M\frac1{\alpha+\beta\varepsilon^{(l)}},
  \label{eq:gc:overN}
\end{equation}
and
\begin{equation}
  \langle U\rangle =
  -
  \left(
    \frac{\partial q}{\partial\beta}
  \right)_{\alpha,\varepsilon,M}=
  \sum_{j=1}^M\frac{\varepsilon^{(l)}}{\alpha+\beta\varepsilon^{(l)}},
  \label{eq:gc:overE}
\end{equation}
where $\varepsilon=\{\varepsilon^{(1)},\ldots,\varepsilon^{(M)}\}$, and $\overline{\varepsilon^{(l)}}=\varepsilon\setminus\varepsilon^{(l)}$. From this point on, unless stated otherwise, we are going to omit the brackets that denote ensemble averaging.
From Eq.~(\ref{eq:RJ}), and utilizing the two conservation laws, we derive the following equation of state
\[
  \alpha N + \beta U = M.
\]
Taking the differential of $q$ and following the relevant calculations we derive
\begin{equation}
  \dif (q+\alpha N+\beta U) = 
  \beta\bigg[\frac\alpha\beta\dif N+ 
  \dif U
  +
  P\dif M
  -
  \sum_{j=1}^J
  R_j\dif\xi_j
  \bigg].
  \label{eq:gc:diff01}
\end{equation}
In Eq.~(\ref{eq:gc:diff01}) we have defined
\begin{equation}
  R_j=
  \sum_{l=1}^M
  n^{(l)}
  \left(
    \frac{\partial\varepsilon^{(l)}}{\partial\xi_j}
  \right)_{\alpha,\beta,M,\overline{\xi_j}}
  \label{eq:gc:R101}
\end{equation}
as the variable that is conjugate to $\xi_j$. In addition the optical pressure~\cite{wu-np2019}
\begin{equation*}
    P =
  \frac1\beta
  \left(
    \frac{\partial q}{\partial M}
  \right)_{\alpha,\beta,\xi},
\end{equation*}
is conjugate to $M$. We compare Eq.~(\ref{eq:gc:diff01}) to the first law of thermodynamics $T\dif S = \delta Q =\dif U -\mu\dif N-\delta W$. From the right hand sides we see that the chemical potential is given by $\mu=-\alpha/\beta$. Furthermore, similarity suggests that we define the work done to the system as 
\begin{equation}
    \delta W =
  -P\dif M
  +
  \sum_{j=1}^J R_j\dif\xi_j.
  \label{eq:deltaW}
\end{equation}
The first term on the right-hand side is the mathematical equivalent to the pressure-volume work with $M$ replacing the volume, while the second term accounts for work due to variations in the system parameters $\xi$.
Comparing the left-hand sides, we derive the following expression for the entropy~\cite{wu-np2019,makri-ol2020}
\begin{equation}
  S = q+\alpha N+\beta U =q+M,
  \label{eq:entropy01}
\end{equation}
and, in addition, we find that
\begin{equation}
  \beta=\frac1T,\quad \alpha=-\frac\mu T.
  \label{eq:alphabeta}
\end{equation}
Thus the equation of state takes the form~\cite{wu-np2019}
\begin{equation}
  U - \mu N = TM. 
  \label{eq:state02}
\end{equation}

\section{First law of thermodynamics and Gibbs-Duhem equation\label{sec:12}}

Utilizing Eqs.~(\ref{eq:gc:diff01}), (\ref{eq:entropy01}) and (\ref{eq:alphabeta}) we conclude that
\begin{equation}
  \dif U = T\dif S -P\dif M +\sum_{j=1}^JR_j\dif\xi_j+\mu\dif N,
  \label{eq:difU}
\end{equation}
which is a manifestation of the 1st law of thermodynamics for multimoded optical systems. We identify the quantities $S$, $M$, $N$ as extensive, and thus the conjugate variables $T$, $p$, $\mu$ are intensive.
We separate the variables $\xi$ into intensive $\xi_{\mathrm{int}}=\{\xi_{\mathrm{int},j}\}$ and extensive $\xi_{\mathrm{ext}}=\{\xi_{\mathrm{ext},j}\}$ segments so that $\xi=\xi_\mathrm{int}\cup\xi_\mathrm{ext}$. Thus, their conjugate variables $R_j$ can also be separated into extensive $R_{\mathrm{ext}}=\{R_{\mathrm{ext},j}\}$ and intensive parts $R_{\mathrm{int}}=\{R_{\mathrm{int},j}\}$, respectively.
We denote by $J_\mathrm{ext}$ the number of extensive variables $\xi_{\mathrm{ext},j}$ and by $J_\mathrm{int}$ the number of intensive variables $\xi_{\mathrm{int},j}$. 
Taking the differential of the Legendre transformation $V = U-\sum_jR_{\mathrm{ext},j}\xi_{\mathrm{int},j}$, and integrating the extensive variables while keeping the intensive variables constant we obtain the following expression for the internal energy
\begin{equation}
  U = TS-PM+\mu N+\sum_{j=1}^{J_\mathrm{ext}}R_{\mathrm{int},j}\xi_{\mathrm{ext},j}.
  \label{eq:U}
\end{equation}

Solving Eq.~(\ref{eq:entropy01}) in terms of $q$ we see that
$q=(TS+\mu N-U)/T$, a formula that can be combined with Eqs.~(\ref{eq:U}) to express the $q$-potential in the form
\begin{equation}
  q = \frac{PM-\sum_{j=1}^{J_\mathrm{ext}}R_{\mathrm{int},j}\xi_{\mathrm{ext},j}}{T}.
  \label{eq:q}
\end{equation}
From Eqs.~(\ref{eq:difU})-(\ref{eq:U}) we derive the Gibbs-Duhem equation
\begin{multline}
  S\dif T-M\dif P-
  \sum_{j=1}^{J_\mathrm{int}}R_{\mathrm{ext},j}\dif\xi_{\mathrm{int},j}+ \\
  \sum_{j=1}^{J_\mathrm{ext}}\xi_{\mathrm{ext},j}\dif R_{\mathrm{int},j}+N\dif\mu = 0
  \label{eq:GibbsDuhem}
\end{multline}
that provides the relationship between variations in the intensive parameters.

\section{Thermodynamics of two-dimensional photonic lattices\label{sec:lattices}}

\subsection{Conservation laws}

As a main example, we focus on two-dimensional photonic lattices consisting of single-mode waveguides that form a rectangular lattice with dimensions $M_1\times M_2$, where $M=M_1M_2$ is the number of modes. In nodal space, we can express the wavefunction as
\[
  |\Psi\rangle= \sum_{m}A_{m}(z)|m\rangle,
\]
where $A_m(z)$ is the $z$-dependent amplitude, $m=(m_1,m_2)$, and $m_j=1,\ldots,M_j$. The two conservation laws of the system~\cite{leder-pr2008} are the total power
\begin{equation}
  N = \sum_m|A_m|^2
  \label{eq:dnls_power}
\end{equation}
and the Hamiltonian
\begin{equation}
  H = -\sum_{m}
  \left[
    A_m^*(\kappa_1\Delta_1A_m+\kappa_2\Delta_2A_m)
    +\frac\gamma2|A_m|^4
  \right],
  \label{eq:dnls_hamiltonian}
\end{equation}
where $\Delta_1A_m=A_{m_1+1,m_2}+A_{m_1-1,m_2}$, and $\Delta_2A_m=A_{m_1,m_2+1}+A_{m_1,m_2-1}$ are the coupling operators between first neighbors along the two transverse directions. The coupling coefficient along the $j$th direction is $\kappa_j$, and we consider interactions of the Kerr type~\cite{boyd-2020} with strength $\gamma$. The discrete nonlinear Schr\"odinger equation 
\begin{equation}
  i\dot A_m +\kappa_1\Delta_1A_m+\kappa_2\Delta_2A_m
  +\gamma |A_m|^2A_m=0,
  \label{eq:dnls}
\end{equation}
where $\dot A_m=\dif A_m/\dif z$, is derived from the Hamiltonian via
$ \dot A_m = \{H,A_m\} $
with Poisson brackets $\{A_m,A_{m'}^*\}=i\delta_{m,m'}$, and $\{A_m,A_{m'}\}=\{A_m^*,A_{m'}^*\}=0$. 
For zero boundary conditions $A_{0,m_2}=A_{M_1+1,m_2}=A_{m_1,0}=A_{m_1,M_2+1}=0$
the lattice supports the following eigenmodes
\begin{equation}
  |l\rangle =
  \sum_m
  \left(
    \prod_{j=1}^2
    \sqrt{\frac{2}{M_j+1}}
    \sin\frac{m_jl_j\pi}{M_j+1}
  \right)
  |m\rangle,
  \label{eq:l}
\end{equation}
with eigenvalues
\begin{equation}
\varepsilon^{(l)}=
  - 2\kappa_1\cos\left(\frac{l_1\pi}{M_1+1}\right)
  - 2\kappa_2\cos\left(\frac{l_2\pi}{M_2+1}\right).
  \label{eq:varepsilon}
\end{equation}
As we can see from Eq.~(\ref{eq:varepsilon}), the system variables $\xi_j$ are the coupling coefficients $\kappa_j$ which are intensive. Their conjugate variables $R_j$ are extensive and are then given by
\[
  R_j
  =-2\sum_ln^{(l)}\cos\frac{2\pi l_j}{M_j+1}
  =-\sum_mA_m^*\Delta_jA_m.
\]
We assume that the cubic nonlinearity is the only source for mode mixing, and is relatively mild in the sense that it does not lead to phase transitions. In such a low intensity limit, the nonlinear contribution to the Hamiltonian is small as compared to the linear terms. Thus, we can write
\begin{equation}
  H = R_1\kappa_1+R_2\kappa_2.
  \label{eq:UofR}
\end{equation}

When the number of modes in rectangular waveguide arrays increases from $M$ to $M'>M$, so that $M_1'\ge M_1$ and $M_2'\ge M_2$, then both the power $N$ and the energy $U$ are conserved. If such a process takes place, say at $z=z_1$, then this can be trivially shown by adding elements with zero amplitude at $z=z_1^+$ in the sums of Eqs.~(\ref{eq:dnls_power})-(\ref{eq:dnls_hamiltonian}). However, when $M_1'<M_1$ or $M_2'<M_2$ then the amount of power and energy that is stored in the waveguides that are eliminated is lost, resulting to a non-reversible process.

The total power does not change if the coupling coefficients are $z$-dependent functions. In contrast, the Hamiltonian or the energy given by Eq.~(\ref{eq:dnls_hamiltonian}) changes as a function of $\kappa_j$ as
\begin{equation}
  \dif H =
  R_1\dif\kappa_1+R_2\dif\kappa_2.
  \label{eq:dotH}
\end{equation}
Utilizing Eqs.~(\ref{eq:UofR}), (\ref{eq:dotH}) we also find that
\begin{equation}
  \kappa_1\dif R_1+\kappa_2\dif R_2=0. 
  \label{eq:difR}
\end{equation}
In Appendix~\ref{sec:app:conserv}, we analyze the dependence of the conservation laws of the discrete nonlinear Sch\"odinger equation from the $z$-dependent coupling coefficients, and derive Eqs.~(\ref{eq:dotH})-(\ref{eq:difR}).

For one-dimensional arrangement of waveguides ($M_2=1$ leading to $R_2=0$) and mild nonlinear effects, we can ignore the nonlinear terms in the Hamiltonian and directly integrate Eq.~(\ref{eq:dotH}). This leads to a Hamiltonian that is proportional to the coupling coefficient
\begin{equation}
  \frac{H(z)}{H(0)}=\frac{\kappa(z)}{\kappa(0)}
  \label{eq:Hz1}
\end{equation}
Equation~(\ref{eq:Hz1}) can be generalized for two-dimensional lattices when the ratio of the coupling coefficients remains constant during propagation, $\kappa_1(z)=\alpha_1\kappa(z)$ and $\kappa_2(z)=\alpha_2\kappa(z)$.

However, when the coupling coefficients $\kappa_1(z)$ and $\kappa_2(z)$ vary independently during propagation, closed form expressions can only be obtained under the additional assumption of thermal equilibrium. In this case, both $R_1$ and $R_2$ should have equal contribution to the Hamiltonian. Physically, this is a condition that warranties that there is no preferable direction in the wave patterns that are generated. For example, stripe soliton solutions whose intensity is maximum along the vertical line $j_1=\mathrm{constant}$ have a strong directional anisotropy $|R_2|\gg|R_1|$. Such a behavior is not expected in thermal equilibrium. From Eq.~(\ref{eq:difR}) and $R=R_1=R_2$ we conclude that $R$ remains constant. Integrating Eq.~(\ref{eq:dotH}) we find that the energy changes during propagation according to
\begin{equation}
  \frac{H(z)}{H(0)}=
  \frac{\kappa_1(z)+\kappa_2(z)}{\kappa_1(0)+\kappa_2(0)}.
  \label{eq:Hz2}
\end{equation}

\subsection{Stress, strain, and extensivity\label{subsec:stress}}

Note that for waveguide arrays the $q$-potential does not depend directly on the number of modes $M$, as we assumed in our analysis, but rather, on the dimensions of the lattice along each direction $M_1$ and $M_2$. Then since $q=q(\alpha,\beta,\kappa_1,\kappa_2,M_1,M_2)$, following calculations similar to those of sections~\ref{sec:11}-\ref{sec:12}, we find that the first law of thermodynamics can be written as 
\begin{multline}
  \dif U = T\dif S
  +\sigma_1M_2\dif M_1
  +\sigma_2M_1\dif M_2
  + \\
  \mu\dif N+R_1\dif\kappa_1+R_2\dif\kappa_2.
  \label{eq:dNLS:difU}
\end{multline}
In Eq.~(\ref{eq:dNLS:difU}) 
\begin{equation}
  \sigma_j=-\frac1{\beta}\frac1{M_{3-j}}
  \left(
    \frac{\partial q}{\partial M_j}
  \right)_{\alpha,\beta,\kappa,M_{3-j}}
\end{equation}
are the
stress terms, whereas $M_j\dif M_{3-j}$ are, the conjugate to the stresses,
strain terms. Since the variables $M_j$ are discrete we can use a difference scheme to express the partial derivative of $q$ with respect to $M_j$ such as
\begin{multline*}
  \left(
    \frac{\partial q}{\partial M_j}
  \right)_{\alpha,\beta,\kappa,M_{3-j}}
  = 
  q(\alpha,\beta,\kappa,M_j+1,M_{3-j})- \\
  q(\alpha,\beta,\kappa,M_j,M_{3-j})
\end{multline*}
In the above expression, note that $M_j$ affects $q$ through both the number of terms that appear in the sum and the function $\varepsilon^{(l)}(M_1,M_2)$ [see Eq.~(\ref{eq:qpot})].

Equation~(\ref{eq:dNLS:difU}) does not explicitly contain the pressure. However, the pressure can be defined as minus the average of the stress terms $\sigma_j$, or
\begin{equation}
  P=-\frac{\sigma_1+\sigma_2}{2}.
  \label{eq:pressure_arrays}
\end{equation}
In addition, we define the stress anisotropy as 
\begin{equation}
  \tau=\frac{\sigma_1-\sigma_2}{2}.
  \label{eq:tau01}
\end{equation}
As a result, utilizing Eqs.~(\ref{eq:pressure_arrays})-(\ref{eq:tau01}) the first law of thermodynamics [Eq.~(\ref{eq:dNLS:difU})] can take the, perhaps, more familiar form
\begin{multline}
  \dif U  = T\dif S-P\dif M + 
  \tau (M_2\dif M_1-M_1\dif M_2)
  + \\
  \mu\dif N+R_1\dif\kappa_1+R_2\dif\kappa_2.
  \label{eq:difU:lattice}
\end{multline}

Interestingly, when $\kappa_1=\kappa_2$ along with $M_1=M_2$ then the $q$-potential depends only on $M$, $q=q(M)$, and $\tau=0$, meaning that the stress is the same along both directions $\sigma_1=\sigma_2$. Thus, we can examine the importance of $\tau$ by studying how well the entropy $S=q(M_1,M_2)+M$ can be approximated by $S(M)$. More importantly, we will also be able to answer the question about whether and when the entropy can be considered as extensive [$S(\lambda U,\lambda N,\lambda M) = \lambda S(U,N,M)$]. From Eq.~(\ref{eq:entropy01}) we see that the entropy is extensive if and only if the $q$ potential is extensive.

\begin{figure*}
\centerline{
  \includegraphics[width=0.8\textwidth]{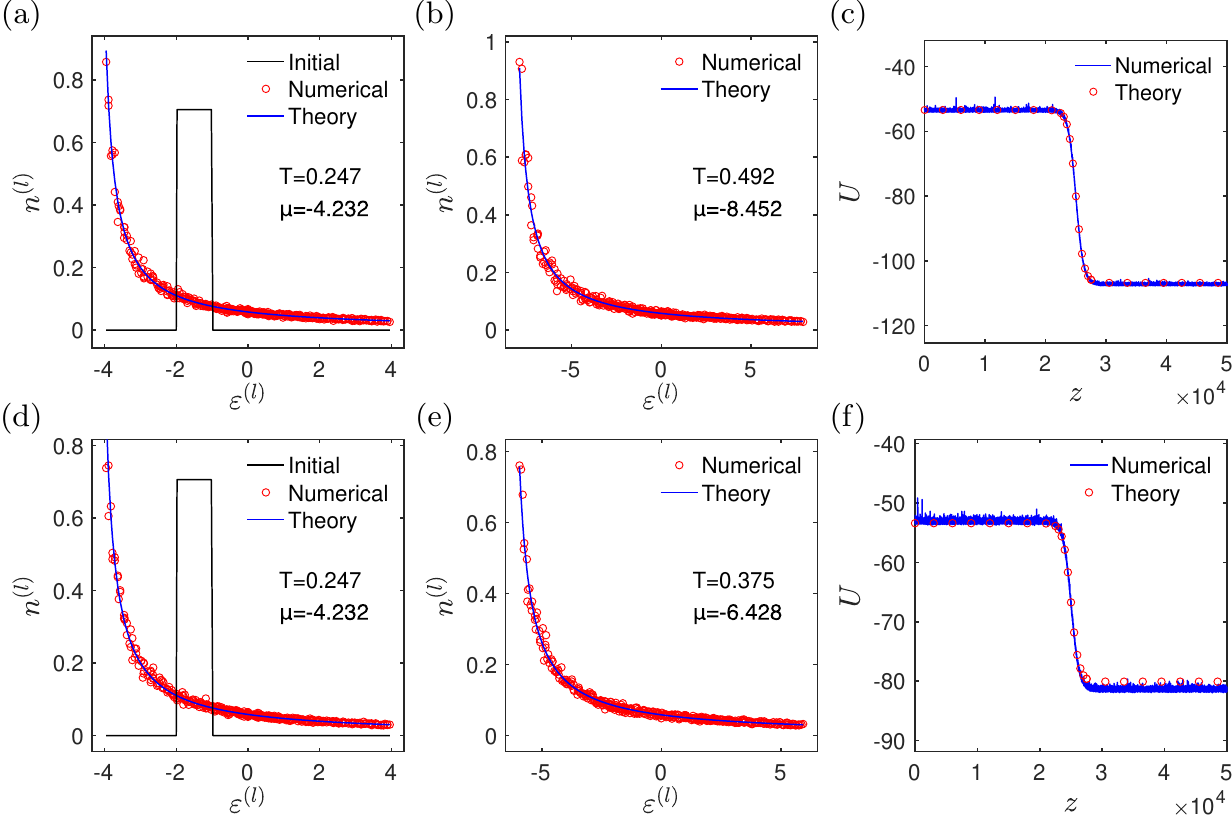}
}
\caption{Numerical results of isentropic processes that take place when the strength of the coupling coefficients changes during propagation in two-dimensional waveguide arrays. In both cases shown in the two rows, the array consists of $20\times20$ waveguides and initially the coupling coefficients are $\kappa_1(0)=\kappa_2(0)=1$. The eigenmodes, lying in the range $-2\le\varepsilon^{(l)}\le-1$, are equally excited with a uniformly distributed random phase, $N=36$, and $U=-53.39$. In (a)-(c) the coupling coefficients $\kappa_1(z)=\kappa_2(z)=\kappa(z)$ smoothly increase to reach the value $\kappa(z_f)=2$ via a sigmoid ($\tanh$) function. In (d)-(f) $\kappa_1(z)$ increases during propagation up to $\kappa_{1}(z_f)=2$ via the same sigmoid ($\tanh$) function, while $\kappa_2(z)=1$. In the first and second column, we see the power distribution of the eigenmodes as a function of the eigenvalue before and after the process, respectively, while in the third column we depict the internal energy as a function of the propagation distance.
  Variations in the values of $\kappa_j$ modify the internal energy according to Eq.~(\ref{eq:Hz2}).  
  \label{fig:1}}
\end{figure*}

Let us first restrict ourselves to the one-dimensional case and assume that $(U',M',N')=\lambda(U,M,N)$. We group the energy levels into cells denoted by $\mathcal C_i$. Assuming that the $i$th cell contains the energy levels in the range $[\varepsilon_{i,l},\varepsilon_{i,r}]$, then the multiplicity of levels $g_i$, is given by
\begin{equation}
  g_i
  =
  \frac{M+1}{\pi}
  \left[
    \arccos\left(-\frac{\varepsilon_{i,r}}{2\kappa}\right)
    -
    \arccos\left(-\frac{\varepsilon_{i,l}}{2\kappa}\right)
  \right].
  \label{eq:gii}
\end{equation}
In cell $\mathcal C_i$ the average energy $\varepsilon_i$, and average power $n_i$ are given by
\[
  g_in_i=\sum_{\varepsilon^{(l)}\in\mathcal C_i}n^{(l)},\quad
  g_in_i\varepsilon_i=\sum_{\varepsilon^{(l)}\in\mathcal C_i}n^{(l)}\varepsilon^{(l)}.
\]
From Eq.~(\ref{eq:gii}), we clearly see that for $M\gg1$ the level multiplicity $g_i$ is proportional to $M$, or $g_i'=\lambda g_i$. From the equation $N'=\lambda N$ we find that $\sum_ig_i(n_i'-n_i)=0$ or $n_i'=n_i$. In addition, from the condition $U'=\lambda U$ we obtain $\sum_ig_in_i(\varepsilon_i'-\varepsilon_i)=0$ and thus $\varepsilon_i'=\varepsilon_i$. Substituting this latter equation to $n_i'=n_i$ results to $T'/(\varepsilon_i-\mu')=T/(\varepsilon_i-\mu)$, an expression that is satisfied when $\mu'=\mu$ and $T'=T$. As a result both the $q$-potential
\[
  q=\sum_ig_i\log\left(\frac{T}{\varepsilon_i-\mu}\right)
\]
and the entropy are extensive. Our numerical results presented in Appendix~\ref{app:ext} are in agreement with these finding (see for example Table~\ref{tab:1}). Let us briefly mention that in our simulations, even for moderate values of $M$, the temperature $T$, the chemical potential $\mu$, and the entropy per mode number $S/M$ remain almost constant.

For two transverse directions, we rely on numerics which show that, besides some limiting cases, the entropy is quasi-extensive, i.e., while not being truly extensive, for a wide range of parameters the deviations from extensivity are small. In all our simulations, we limit ourselves to the case $U<0$ (and thus $T>0$). However, taking advantage of the bijective mapping between $(N,U,T,\mu)$ and $(N,-U,-T,-\mu)$, our results can also be applied for $U>0$ and thus $T<0$ (see Appendix~\ref{sec:negtemp} for details).
Specifically, we vary $M_1$ and $M_2$ while keeping $M$ constant and, subsequently, we change the value of $M$ and follow the same procedure. Deviations from the extensive nature of the entropy start to take place as we approach the one-dimensional limit, where the size of the array along one direction is small. Such deviations decrease as we increase the power or decrease the energy. Anisotropy in the coupling coefficients can also mildly affect the extensivity of the entropy. 
As we have shown, such variations from the extensive character of the entropy, give rise to stress anisotropy $\tau$ which, in general, is small. 
In Tables~\ref{tab:2}-\ref{tab:5} we depict $S/M$, $T$ and $\mu$ as a function of the lattice dimensions $M_1$ and $M_2$ for different values of the coupling coefficients $\kappa_1$ and $\kappa_2$.

\begin{figure*}
\centerline{
  \includegraphics[width=0.8\textwidth]{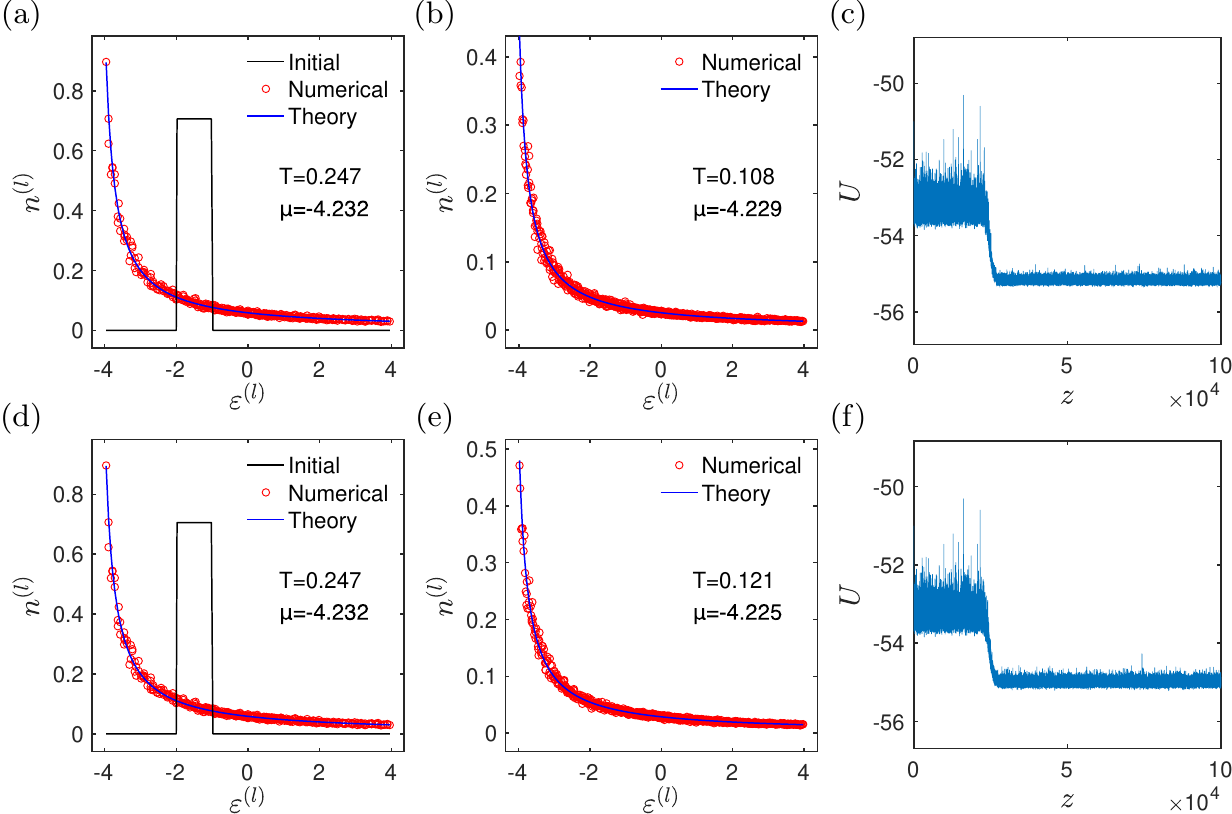}
}
\caption{Numerical results of Joule-type expansions associated with an increase in the number of modes in two-dimensional waveguide arrays. The figure arrangement and initial parameters are the same as in Fig.~\ref{fig:1}. In the first row, the number of modes increases during propagation up to $M_1=M_2=30$, so that at every propagation distance the lattice remains square. In the second row, only the dimension of the lattice along the first direction increases during propagation up to $M_1=40$. In both examples the number of waveguides is determined by a sigmoid ($\tanh$) function rounded to the closest integer.\label{fig:2}}
\end{figure*}

\subsection{Thermodynamic processes}

We are going to investigate the processes that arise due to variations in the parameters of the waveguide arrays both theoretically and numerically. In our simulations, a part of the spectrum is initially excited with equal amplitude and uniformly distributed phase, but is allowed to reach thermal equilibrium before the process take place. Ensemble averaging is obtained by averaging in $z$,
\[
  \langle n^{(l)}\rangle =\frac1{z_2-z_1}\int_{z_1}^{z_2}n^{(l)}(z)\dif z.
\]
For simplicity, in our simulations we consider lattices of comparable dimensions $M_1$ and $M_2$. Thus effects due to stress anisotropy are small and are ignored for the rest of this work. 

Let us first study processes, which take place when the coupling coefficients change during propagation in two-dimensional rectangular lattices. Such processes were examined in the one-dimensional case in~\cite{wu-np2019} and were found to be isentropic and useful in achieving optical refrigeration. Since the number of modes and the power remain constant, the 1st law of thermodynamics given by Eq.~(\ref{eq:difU:lattice}) takes the form $\dif U = T\dif S+R_1\dif\kappa_1+R_2\dif\kappa_2$. Combining the previous equation with Eq.~(\ref{eq:dotH}) we find that $\dif S=0$. Thus, variations in the coupling coefficients in two-dimensional lattices conserve the entropy as well as the $q$-potential. This procedure can be extended in the case of lattices with higher dimensionality, to show that they are also isentropic.

Let us first consider the case where the coupling coefficients along the two directions are identical
$\kappa_1(z)=\kappa_2(z)=\kappa(z)$ and smoothly change from $\kappa(0)=\kappa$ to $\kappa(z_f)=\kappa'=a\kappa$, where $z_f$ is the simulation distance. Since the energies scale linearly with $\kappa(z)$ we have $(\varepsilon^{(l)})'=a\varepsilon^{(l)}$, and utilizing the conservation of power $N=N'$, we find that $T'=a T$ and $\mu'=a\mu$. During this process the pressure changes to $P'=a P$. As for the energy, integrating 
$\dif U = \dif W = R_1\dif\kappa_1+R_2\dif\kappa_2= U\dif\kappa/\kappa$, we obtain that $U'=a U$. This latter equation is in agreement with Eq.~(\ref{eq:Hz1}), which was derived using the Hamiltonian structure of the lattice. In Figs.~\ref{fig:1}(a)-(c) we see numerical results in the case where $\kappa_1(z)=\kappa_2(z)=\kappa(z)$ with $\kappa(0)=1$ and $\kappa(z_f)=2$.
The power modal occupancies shown in Fig.~\ref{fig:1}(a)-(b) are in excellent agreement with the numerical results before and after the process. 
Finally in Fig.~\ref{fig:1}(c) we see the variation of the internal energy of the array as a function of the propagation distance.

In the more general case where both coupling coefficients vary independently during propagation, we have shown that under thermal equilibrium Eq.~(\ref{eq:Hz2}), that relates the energy to the coupling coefficients, is satisfied. In our simulations, (i) the system is allowed to reach thermal equilibrium before the process takes place and, (ii) during the process the variations of the coupling coefficients are slow enough so that the system remains close to thermal equilibrium. An example is shown in Fig.~\ref{fig:1}(d)-(f), where $\kappa_1$ increases during propagation while $\kappa_2$ remains constant. The comparison between theory and numerics is very good both in the power distributions, as well as in the comparison of Eq.~(\ref{eq:Hz2}) with the variation in the values of the internal energy.

We also examine the process during which the number of waveguides increases from $M$ to $M'=a M$ ($a>1$).
Such a process can be considered as the equivalent of Joule expansion (or free expansion) for photonic lattices.
We separate the energy levels into cells $\mathcal C_i$  and follow an analysis  similar to the one utilized in Section~\ref{subsec:stress}. In particular, since both $N$ and $U$ are conserved, we find that, after the process takes place, the multiplicity of energy levels in cell $\mathcal C_i$ increases by a factor of $a$ or $g_i'=a g_i$.
Following the calculations it can be shown that the chemical potential $\mu$ remains invariant, while both the temperature $T'=T/a$ and the average power per mode $n_i'=n_i/a$ in cell $\mathcal C_i$ decrease by a factor of $a$. The $q$-potential after the process becomes
\[
  q' = \sum_ig_i'\log n_i'=a q -a M\log a,
\]
from which we can compute the entropy $S' = a S-a M\log a$ and the pressure $P'M'=PM-MT\log a$.
The 1st law of thermodynamics leads to the differential $T\dif S=P\dif M$ while the Gibbs-Duhem equation results to $S\dif T=M\dif P$. Note that for such processes $TS-MP=U-\mu N = TM$ is a constant. Thus, as the number of waveguides increases, the temperature decreases in a fashion that is inversely proportional to the number of waveguides $T\propto1/M$.

In Fig.~\ref{fig:2} we depict numerical results for such Joule-type expansions. In the first simulation shown in~\ref{fig:2}(a)-(c) the number of modes along each direction mutually increases during propagation so that $M_1(z)=M_2(z)$ using a sigmoid $\tanh$ function with rounding to the closest integer. In Fig.~\ref{fig:2}(a)-(b), we compare the theoretical with the numerical power modal occupancies before and after the process. In Fig.~\ref{fig:2}(c), we note there is a small amount of decrease in the internal energy during the process. This happens because as the number of modes increases, the power is spread into more waveguides, which reduces the intensity of light and thus the absolute value of the nonlinear term of the Hamiltonian.
In Fig.~\ref{fig:2}(d)-(f) we see similar results, but now the number of waveguides increases only along the first direction.

As already mentioned, such a process is non-reversible. If, during propagation, we decrease the number of waveguides ($a<1$), then both the power and the energy stored in these waveguides is lost. Thus the results obtained in this section hold only in the case of expansion.

\section{Conclusions}

In conclusion, we have worked along the direction of generating a complete thermodynamic theory of nonlinear optics. In particular, we have introduced the optical analogues of the 1st law of thermodynamics, the work done to the system, stress and strain terms to account for anisotropy, analyzed the role of the system parameters, and examined different types of processes (isentropic and Joule-type expansions). Besides photonic lattices, which is the particular system that we apply our theory, our results can also be utilized in a variety of multimoded optical system, such as optical fibers and coupled microresonators, as well as in different physical settings such as Bose-Einstein condensates. 

\appendix

\section{Relations between the thermodynamic parameters in waveguide arrays\label{sec:negtemp}}

In our simulations in waveguide arrays we focus in the case of negative internal energies $U$ that give rise to positive temperatures $T$. Using coupled mode theory for two-dimensional rectangular waveguide arrays, the eigenmodes $|l\rangle$ and eigenvalues $\varepsilon^{(l)}$ are given by Eqs.~(\ref{eq:l})-(\ref{eq:varepsilon}), where $M_j$ is the length of the lattice and $\kappa_j$ is the coupling coefficient in the $j$th direction, $M=M_1M_2$ is the number of supported modes, and $l=(l_1,l_2)$. We note that the eigenvalues have the symmetry
\begin{equation}
  \varepsilon^{(M_1-l_1+1,M_2-l_2+1)}=-
  \varepsilon^{(l_1,l_2)}
  \label{eq:epsilonrel}
\end{equation}
or, using vector notation $\varepsilon^{(M_v-l+c)}=-\varepsilon^{(l)}$, where $M_v=(M_1,M_2)$, and $c=(1,1)$. Furthermore, the energies lie in the range $[\varepsilon^{(1,1)},\varepsilon^{(M_1,M_2)}]$ with
\begin{equation*}
  -2(\kappa_1+\kappa_2)<\varepsilon^{(1,1)}<0<
  \varepsilon^{(M_1,M_2)}<2(\kappa_1+\kappa_2).
\end{equation*}
The eigeneneries have the limiting values $\varepsilon^{(1,1)}\rightarrow-2(\kappa_1+\kappa_2)$ and $\varepsilon^{(M_1,M_2)}\rightarrow2(\kappa_1+\kappa_2)$ when both $M_1$ and $M_2$ go to infinity. Since the power occupation numbers 
\begin{equation}
  n^{(l)}=\frac{T}{\varepsilon^{(l)}-\mu}
  \label{eq:nl}
\end{equation}
are all positive, selecting a positive temperature $T>0$ result to $\mu<\varepsilon^{(1,1)}<0$. Thus, we note that lower energies have higher occupation numbers. The two conserved quantities are the total or internal energy [Eq.~(\ref{eq:U0})]
and the total power [Eq.~(\ref{eq:N})].
Utilizing Eq.~(\ref{eq:epsilonrel}) can express the internal energy as 
\begin{equation}
  U =
  \sum_{l_-}^{}\varepsilon^{(l)}
  [n^{(l)}-n^{(M_v-l+c)}],
  \label{eq:U1}
\end{equation}
where the summation is over the indices with negative energies $l_-=\{l:\varepsilon^{(l)}<0\}$. Since the lower energies have higher occupation numbers and $\varepsilon^{(l)}<0<\varepsilon^{(M_v-l+c)}$ we find that $n^{(l)}>n^{(M-l+1)}$, and thus the internal energy is negative $U<0$. 

We consider a system in thermal equilibrium with $T>0$, $\mu<0$, $U<0$, power $N$, and power occupation numbers $n^{(l)}$. We would like to find the thermodynamic parameters of a system with the same energy levels and occupation numbers that satisfy
\begin{equation}
  (n^{(l)})'=n^{(M_v-l+c)}.
  \label{eq:app:nlp}
\end{equation}
Starting from Eq.~(\ref{eq:nl}) and utilizing Eqs.~(\ref{eq:epsilonrel}), (\ref{eq:app:nlp}) we conclude that 
\[
  (n^{(l)})'=\frac{T'}{\varepsilon^{(l)}-\mu'},
\]
where $T'=-T$ and $\mu'=-\mu$. In addition, from Eq.~(\ref{eq:U0}) we obtain $U'=-U$. Both the $q$-potential $q=\sum_{l=1}^M\log n^{(l)}$ and the entropy $S=q+M$ remain invariant ($q'=q$ and $S'=S$) while the pressure $P=qM/T$ changes sign ($P'=-P$). Thus, for the system with occupation numbers given by Eq.~(\ref{eq:app:nlp}) we have $U'=-U>0$, $T'=-T<0$, $\mu'=-\mu>0$, and $P'=-P$. 

The relation between the signs of $T$ and $U$ is also obvious from the definition of the temperature $1/T=(\partial S/\partial U)_{M,N,\kappa}$, by noting the that number of microstates (and thus the entropy) is maximized in the middle of the band where $U=0$. 

The results presented here are not limited to rectangular arrangements of waveguides but can be applied to photonic lattices with different geometries, or even different classes of multimoded optical systems, as long as the energy spectrum satisfies a bijective relation between opposite energies ($\varepsilon$ and $-\varepsilon$).

\section{Extensive character of the entropy: numerical results\label{app:ext}}

In two-dimensional waveguide arrays, the entropy depends on the lattice dimensions ($M_1$ and $M_2$) rather than the number of modes $M=M_1M_2$. A question that naturally arises is, up to what degree, we can consider the entropy as an extensive function of $U$, $M$, $N$ or $S(\lambda U,\lambda M,\lambda N)=\lambda S(U,M,N)$. Here, we are going to examine numerically the variations in the entropy per number of modes $S_M=S/M$, for different values of $(M_1,M_2)$ and different coupling coefficients $(\kappa_1,\kappa_2)$. We restrict ourselves to $T>0$ (or equivalently $U<0$). These results can be trivially generalized for negative temperatures (or positive internal energies) by applying the transformations of Appendix~\ref{sec:negtemp}. The numerical results presented in this appendix are obtained by numerically solving Eqs.~(\ref{eq:gc:overN}), (\ref{eq:gc:overE}), where $U$ and $N$ are related through Eq.~(\ref{eq:state02}), and the eigenvalues are given by Eq.~(\ref{eq:varepsilon}). The process is described in~\cite{parto-ol2019}.

\begin{table}
  \centerline{
    \begin{tabular}{rr|rrr} \toprule \toprule
      \multicolumn{2}{c}{$\kappa=1$}\vline & \multicolumn{3}{c}{$\frac NM=\frac{1}{50}$, $\frac UM=-\frac{3}{800}$} \\\midrule
      $M_1$ & $M_2$ & $T$ & $\mu$ & $S/M$ \\\midrule
      $2$ & $1$ & $0.103$ & $-5.33$ & $-2.930$ \\
      $5$ & $1$ & $0.169$ & $-8.63$ & $-2.923$ \\
      $10$ & $1$ & $0.190$ & $-9.70$ & $-2.922$ \\
      $20$ & $1$ & $0.201$ & $-10.23$ & $-2.921$ \\
      $25$ & $1$ & $0.203$ & $-10.33$ & $-2.921$ \\
      $50$ & $1$ & $0.207$ & $-10.55$ & $-2.921$ \\
      $100$ & $1$ & $0.209$ & $-10.65$ & $-2.921$ \\
      $200$ & $1$ & $0.210$ & $-10.70$ & $-2.921$ \\
      $400$ & $1$ & $0.211$ & $-10.73$ & $-2.921$ \\
      $800$ & $1$ & $0.211$ & $-10.74$ & $-2.921$ \\
      $1600$ & $1$ & $0.211$ & $-10.75$ & $-2.921$ \\
      $3200$ & $1$ & $0.211$ & $-10.76$ & $-2.921$ \\
      $6400$ & $1$ & $0.211$ & $-10.76$ & $-2.921$ \\
      \bottomrule\bottomrule
    \end{tabular}
  }
  \caption{Temperature $T$, chemical potential $\mu$, and entropy per mode number $S/M$ as a function of the number of modes $M=M_1$ for one-dimensional waveguide arrays ($M_2=1$) with $\kappa=1$. The power per mode number and energy per mode number are $N/M=1/50$, $U/M=-3/800$.\label{tab:1}}
\end{table}
\begin{table*}
  \setlength{\tabcolsep}{4pt}
  \centerline{
    \small
    \begin{tabular}{rr|rrr|rrr|rrr|rrr} \toprule\toprule
            \multicolumn{2}{c}{$\kappa_1=\kappa_2=1$}\vline & \multicolumn{3}{c}{$\frac NM=\frac{1}{400}$, $\frac UM=-\frac{3}{800}$}\vline  &\multicolumn{3}{c}{$\frac NM=\frac{1}{50}$, $\frac UM=-\frac{3}{800}$}\vline & \multicolumn{3}{c}{$\frac NM=\frac{1}{400}$, $\frac UM=-\frac{1}{1600}$}\vline & \multicolumn{3}{c}{$\frac NM=\frac{1}{50}$, $\frac UM=-\frac{1}{1600}$} \\\midrule
      $M_1$ & $M_2$ & $T$ & $\mu$ & $S/M$ & $T$ & $\mu$ & $S/M$ & $T$ & $\mu$ & $S/M$& $T$ & $\mu$ & $S/M$\\\midrule
      $100$ & $1$ & $0.0014$ & $-2.08$ & $-5.827$ & $0.209$ & $-10.65$ & $-2.921$& $0.019$ & $-8.05$ & $-5.007$& $0.401$ & $-35.67$ & $-3.488$\\
      $50$ & $2$ & $0.0045$ & $-3.31$ & $-5.379$ & $0.316$ & $-16.00$ & $-2.918$& $0.030$ & $-12.12$ & $-5.002$& $0.599$ & $-53.34$ & $-3.488$\\
      $25$ & $4$ & $0.0061$ & $-3.92$ & $-5.306$ & $0.366$ & $-18.47$ & $-2.917$& $0.034$ & $-13.99$ & $-5.001$& $0.693$ & $-61.63$ & $-3.488$\\
      $20$ & $5$ & $0.0063$ & $-4.01$ & $-5.297$ & $0.374$ & $-18.90$ & $-2.917$ & $0.035$ & $-14.31$ & $-5.000$& $0.709$ & $-63.07$ & $-3.488$\\
      $10$ & $10$ & $0.0065$ & $-4.10$ & $-5.288$  & $0.385$ & $-19.44$ & $-2.917$& $0.036$ & $-14.72$ & $-5.000$& $0.729$ & $-64.87$ & $-3.488$\\
      $400$ & $1$ & $0.0015$ & $-2.08$ & $-5.820$ & $0.211$ & $-10.73$ & $-2.921$ & $0.020$ & $-8.11$ & $-5.007$& $0.404$ & $-35.94$ & $-3.488$\\
      $200$ & $2$ & $0.0046$ & $-3.33$ & $-5.375$ & $0.319$ & $-16.15$ & $-2.918$ & $0.030$ & $-12.24$ & $-5.002$& $0.606$ & $-53.88$ & $-3.488$ \\
      $100$ & $4$ & $0.0061$ & $-3.95$ & $-5.301$ & $0.372$ & $-18.79$ & $-2.917$ & $0.035$ & $-14.23$ & $-5.000$& $0.705$ & $-62.71$ & $-3.488$ \\
      $50$ & $8$ & $0.0066$ & $-4.16$ & $-5.281$  & $0.397$ & $-20.02$ & $-2.917$  & $0.037$ & $-15.15$ & $-5.000$& $0.751$ & $-66.85$ & $-3.488$ \\
      $25$ & $16$ & $0.0068$ & $-4.22$ & $-5.274$ & $0.406$ & $-20.48$ & $-2.917$ & $0.038$ & $-15.49$ & $-5.000$& $0.769$ & $-68.38$ & $-3.488$ \\
      $20$ & $20$ & $0.0068$ & $-4.22$ & $-5.274$ & $0.406$ & $-20.50$ & $-2.917$& $0.038$ & $-15.51$ & $-5.000$& $0.770$ & $-68.47$ & $-3.488$\\
      $1600$ & $1$ & $0.0015$ & $-2.08$ & $-5.819$ & $0.211$ & $-10.75$ & $-2.921$& $0.020$ & $-8.12$ & $-5.007$ & $0.404$ & $-36.01$ & $-3.488$\\
      $800$ & $2$ & $0.0046$ & $-3.33$ & $-5.375$ & $0.320$ & $-16.19$ & $-2.918$& $0.030$ & $-12.27$ & $-5.002$& $0.607$ & $-54.02$ & $-3.488$\\
      $400$ & $4$ & $0.0062$ & $-3.96$ & $-5.300$ & $0.374$ & $-18.87$ & $-2.917$& $0.035$ & $-14.29$ & $-5.000$& $0.708$ & $-62.98$ & $-3.488$\\
      $200$ & $8$ & $0.0067$ & $-4.17$ & $-5.279$ & $0.400$ & $-20.18$ & $-2.917$& $0.038$ & $-15.27$ & $-5.000$& $0.758$ & $-67.39$ & $-3.488$\\
      $100$ & $16$ & $0.0069$ & $-4.24$ & $-5.271$ & $0.412$ & $-20.80$ & $-2.917$& $0.039$ & $-15.73$ & $-5.000$& $0.781$ & $-69.46$ & $-3.488$\\
      $50$ & $32$ & $0.0069$ & $-4.27$ & $-5.268$ & $0.417$ & $-21.02$ & $-2.917$& $0.039$ & $-15.90$ & $-4.999$& $0.789$ & $-70.22$ & $-3.488$\\
      $40$ & $40$ & $0.0069$ & $-4.27$ & $-5.268$  & $0.417$ & $-21.03$ & $-2.917$& $0.039$ & $-15.91$ & $-4.999$ & $0.790$ & $-70.27$ & $-3.488$ \\
      \bottomrule\bottomrule
    \end{tabular}
  }
  \caption{Temperature $T$, chemical potential $\mu$, and entropy per mode number $S/M$ as a function of the number of waveguides along each direction $M_1$, $M_2$ for two-dimensional rectangular waveguide arrays with $\kappa_1=\kappa_2=1$. The power per mode number and energy per mode number are $N/M=1/400$, $U/M=-3/800$ in columns 3-5, $N/M=1/50$, $U/M=-3/800$, in columns 6-8, $N/M=1/400$, $U/M=-1/1600$ in columns 9-11, and $N/M=1/50$, $U/M=-1/1600$ in columns 12-14.\label{tab:2}}
\end{table*}


\begin{table*}
  \setlength{\tabcolsep}{3pt}
  \centerline{
    \begin{tabular}{rr|rrr|rrr|rrr|rrr} \toprule\toprule
                  \multicolumn{2}{c}{$\kappa_1=1,\kappa_2=2$}\vline & \multicolumn{3}{c}{$\frac NM=\frac{1}{400}$, $\frac UM=-\frac{3}{800}$}\vline  &\multicolumn{3}{c}{$\frac NM=\frac{1}{50}$, $\frac UM=-\frac{3}{800}$}\vline & \multicolumn{3}{c}{$\frac NM=\frac{1}{400}$, $\frac UM=-\frac{1}{1600}$}\vline & \multicolumn{3}{c}{$\frac NM=\frac{1}{50}$, $\frac UM=-\frac{1}{1600}$} \\\midrule
      $M_1$ & $M_2$ & $T$ & $\mu$ & $S/M$ & $T$ & $\mu$ & $S/M$ & $T$ & $\mu$ & $S/M$ & $T$ & $\mu$ & $S/M$ \\\midrule
      $400$ & $1$ & $0.001$ & $-2.08$ & $-5.820$ & $0.211$ & $-10.73$ & $-2.921$ & $0.020$ & $-8.11$ & $-5.007$ & $0.404$ & $-35.94$ & $-3.488$ \\
      $200$ & $2$ & $0.009$ & $-5.25$ & $-5.184$ & $0.639$ & $-32.12$ & $-2.915$ & $0.060$ & $-24.19$ & $-4.997$ & $1.213$ & $-107.87$ & $-3.488$\\
      $100$ & $4$ & $0.014$ & $-6.91$ & $-5.131$  & $0.852$ & $-42.76$ & $-2.914$ & $0.080$ & $-32.19$ & $-4.995$ & $1.616$ & $-143.70$ & $-3.488$\\
      $50$ & $8$ & $0.015$ & $-7.53$ & $-5.116$ & $0.956$ & $-47.98$ & $-2.914$ & $0.090$ & $-36.10$ & $-4.995$ & $1.814$ & $-161.34$ & $-3.488$ \\
      $40$ & $10$ & $0.015$ & $-7.64$ & $-5.114$ & $0.976$ & $-48.99$ & $-2.914$ & $0.092$ & $-36.86$ & $-4.995$ & $1.853$ & $-164.76$ & $-3.488$\\
      $20$ & $20$ & $0.016$ & $-7.83$ & $-5.110$  & $1.013$ & $-50.85$ & $-2.914$ & $0.095$ & $-38.25$ & $-4.995$ & $1.924$ & $-171.06$ & $-3.488$\\
      $10$ & $40$ & $0.016$ & $-7.86$ & $-5.109$ & $1.024$ & $-51.38$ & $-2.914$ & $0.096$ & $-38.64$ & $-4.995$ & $1.944$ & $-172.85$ & $-3.488$\\
      $5$ & $80$ & $0.016$ & $-7.74$ & $-5.111$ & $1.013$ & $-50.84$ & $-2.914$ & $0.095$ & $-38.23$ & $-4.995$ & $1.924$ & $-171.05$ & $-3.488$\\
      $4$ & $100$ & $0.015$ & $-7.66$ & $-5.112$ & $1.004$ & $-50.41$ & $-2.914$ & $0.094$ & $-37.91$ & $-4.995$ & $1.907$ & $-169.61$ & $-3.488$\\
      $2$ & $200$ & $0.014$ & $-7.17$ & $-5.120$ & $0.955$ & $-47.94$ & $-2.914$ & $0.089$ & $-36.04$ & $-4.995$ & $1.814$ & $-161.32$ & $-3.488$\\
      $1$ & $400$ & $0.011$ & $-6.07$ & $-5.143$ & $0.849$ & $-42.65$ & $-2.914$ & $0.079$ & $-32.05$ & $-4.995$ & $1.616$ & $-143.67$ & $-3.488$\\
      $1600$ & $1$ & $0.001$ & $-2.08$ & $-5.819$ & $0.211$ & $-10.75$ & $-2.921$ & $0.020$ & $-8.12$ & $-5.007$ & $0.404$ & $-36.01$ & $-3.488$ \\
      $800$ & $2$ & $0.009$ & $-5.26$ & $-5.184$ & $0.640$ & $-32.16$ & $-2.915$ & $0.060$ & $-24.22$ & $-4.997$ & $1.214$ & $-108.01$ & $-3.488$\\
      $320$ & $5$ & $0.014$ & $-7.19$ & $-5.124$ & $0.896$ & $-44.97$ & $-2.914$ & $0.084$ & $-33.84$ & $-4.995$ & $1.700$ & $-151.15$ & $-3.488$\\
      $160$ & $10$ & $0.015$ & $-7.67$ & $-5.113$ & $0.980$ & $-49.19$ & $-2.914$ & $0.092$ & $-37.01$ & $-4.995$ & $1.860$ & $-165.43$ & $-3.488$\\
      $80$ & $20$ & $0.016$ & $-7.89$ & $-5.109$ & $1.021$ & $-51.26$ & $-2.914$ & $0.096$ & $-38.55$ & $-4.995$ & $1.939$ & $-172.41$ & $-3.488$\\
      $40$ & $40$ & $0.016$ & $-7.98$ & $-5.107$ & $1.040$ & $-52.19$ & $-2.914$ & $0.097$ & $-39.25$ & $-4.995$ & $1.974$ & $-175.55$ & $-3.488$\\
      $20$ & $80$ & $0.016$ & $-8.00$ & $-5.107$ & $1.045$ & $-52.45$ & $-2.914$ & $0.098$ & $-39.44$ & $-4.995$ & $1.984$ & $-176.45$ & $-3.488$\\
      $10$ & $160$ & $0.016$ & $-7.94$ & $-5.107$ & $1.040$ & $-52.18$ & $-2.914$ & $0.097$ & $-39.24$ & $-4.995$ & $1.974$ & $-175.55$ & $-3.488$\\
      $5$ & $320$ & $0.016$ & $-7.78$ & $-5.110$ & $1.021$ & $-51.24$ & $-2.914$ & $0.096$ & $-38.53$ & $-4.995$ & $1.939$ & $-172.40$ & $-3.488$\\
      $2$ & $800$ & $0.014$ & $-7.19$ & $-5.120$ & $0.958$ & $-48.10$ & $-2.914$ & $0.090$ & $-36.16$ & $-4.995$ & $1.820$ & $-161.86$ & $-3.488$\\
      $1$ & $1600$ & $0.011$ & $-6.08$ & $-5.143$  & $0.851$ & $-42.73$ & $-2.914$ & $0.080$ & $-32.11$ & $-4.995$ & $1.619$ & $-143.94$ & $-3.488$\\
      \bottomrule\bottomrule
    \end{tabular}
  }
    \caption{Temperature $T$, chemical potential $\mu$, and entropy per mode number $S/M$ as a function of the number of waveguides along each direction $M_1$, $M_2$ for two-dimensional rectangular waveguide arrays with $\kappa_1=1$, $\kappa_2=2$. The power per mode number and energy per mode number are $N/M=1/400$, $U/M=-3/800$ in columns 3-5, $N/M=1/50$, $U/M=-3/800$, in columns 6-8, $N/M=1/400$, $U/M=-1/1600$ in columns 9-11, and $N/M=1/50$, $U/M=-1/1600$ in columns 12-14.\label{tab:3}}
\end{table*}

\begin{table*}
  \setlength{\tabcolsep}{3pt}
  \centerline{
    \begin{tabular}{rr|rrr|rrr|rrr|rrr} \toprule\toprule
      \multicolumn{2}{c}{$\kappa_1=1,\kappa_2=4$}\vline & \multicolumn{3}{c}{$\frac NM=\frac{1}{400}$, $\frac UM=-\frac{3}{800}$}\vline  &\multicolumn{3}{c}{$\frac NM=\frac{1}{50}$, $\frac UM=-\frac{3}{800}$}\vline & \multicolumn{3}{c}{$\frac NM=\frac{1}{400}$, $\frac UM=-\frac{1}{1600}$}\vline & \multicolumn{3}{c}{$\frac NM=\frac{1}{50}$, $\frac UM=-\frac{1}{1600}$} \\\midrule
      $M_1$ & $M_2$ & $T$ & $\mu$ & $S/M$ & $T$ & $\mu$ & $S/M$ & $T$ & $\mu$ & $S/M$ & $T$ & $\mu$ & $S/M$\\\midrule
      $400$ & $1$ & $0.001$ & $-2.08$ & $-5.820$ & $0.211$ & $-10.73$ & $-2.921$ & $0.020$ & $-8.11$ & $-5.007$ & $0.404$ & $-35.94$ & $-3.488$\\
      $200$ & $2$ & $0.028$ & $-12.62$ & $-5.056$ & $1.917$ & $-96.02$ & $-2.913$ & $0.180$ & $-72.06$ & $-4.993$ & $3.643$ & $-323.84$ & $-3.487$\\
      $100$ & $4$ & $0.042$ & $-18.47$ & $-5.035$ & $2.770$ & $-138.70$ & $-2.913$ & $0.260$ & $-104.11$ & $-4.993$ & $5.261$ & $-467.68$ & $-3.487$\\
      $50$ & $8$ & $0.049$ & $-21.08$ & $-5.029$ & $3.195$ & $-159.92$ & $-2.913$ & $0.299$ & $-120.02$ & $-4.993$ & $6.067$ & $-539.32$ & $-3.487$\\
      $40$ & $10$ & $0.050$ & $-21.59$ & $-5.028$ & $3.279$ & $-164.13$ & $-2.913$ & $0.307$ & $-123.18$ & $-4.992$ & $6.227$ & $-553.54$ & $-3.487$\\
      $20$ & $20$ & $0.053$ & $-22.58$ & $-5.027$ & $3.444$ & $-172.40$ & $-2.913$ & $0.323$ & $-129.37$ & $-4.992$ & $6.541$ & $-581.44$ & $-3.487$\\
      $10$ & $40$ & $0.054$ & $-23.01$ & $-5.026$ & $3.519$ & $-176.12$ & $-2.913$ & $0.330$ & $-132.17$ & $-4.992$ & $6.682$ & $-594.04$ & $-3.487$ \\
      $5$ & $80$ & $0.054$ & $-23.11$ & $-5.026$ & $3.540$ & $-177.19$ & $-2.913$ & $0.332$ & $-132.96$ & $-4.992$ & $6.723$ & $-597.64$ & $-3.487$\\
      $4$ & $100$ & $0.054$ & $-23.08$ & $-5.026$ & $3.538$ & $-177.08$ & $-2.913$ & $0.332$ & $-132.88$ & $-4.992$ & $6.719$ & $-597.28$ & $-3.487$\\
      $2$ & $200$ & $0.053$ & $-22.79$ & $-5.026$ & $3.501$ & $-175.26$ & $-2.913$ & $0.328$ & $-131.51$ & $-4.992$ & $6.650$ & $-591.15$ & $-3.487$\\
      $1$ & $400$ & $0.051$ & $-22.03$ & $-5.027$ & $3.403$ & $-170.33$ & $-2.913$ & $0.319$ & $-127.81$ & $-4.992$ & $6.463$ & $-574.59$ & $-3.487$\\
      $1600$ & $1$ & $0.001$ & $-2.08$ & $-5.819$ & $0.211$ & $-10.75$ & $-2.921$ & $0.020$ & $-8.12$ & $-5.007$ & $0.404$ & $-36.01$ & $-3.488$\\
      $800$ & $2$ & $0.028$ & $-12.63$ & $-5.056$ & $1.917$ & $-96.06$ & $-2.913$ & $0.180$ & $-72.09$ & $-4.993$ & $3.644$ & $-323.98$ & $-3.487$\\
      $320$ & $5$ & $0.045$ & $-19.54$ & $-5.033$ & $2.942$ & $-147.31$ & $-2.913$ & $0.276$ & $-110.56$ & $-4.993$ & $5.588$ & $-496.73$ & $-3.487$\\
      $160$ & $10$ & $0.050$ & $-21.62$ & $-5.028$ & $3.283$ & $-164.33$ & $-2.913$ & $0.308$ & $-123.33$ & $-4.992$ & $6.234$ & $-554.21$ & $-3.487$\\
      $80$ & $20$ & $0.053$ & $-22.63$ & $-5.027$ & $3.452$ & $-172.80$ & $-2.913$ & $0.324$ & $-129.67$ & $-4.992$ & $6.556$ & $-582.79$ & $-3.487$\\
      $40$ & $40$ & $0.054$ & $-23.13$ & $-5.026$ & $3.535$ & $-176.93$ & $-2.913$ & $0.331$ & $-132.77$ & $-4.992$ & $6.713$ & $-596.74$ & $-3.487$\\
      $20$ & $80$ & $0.055$ & $-23.34$ & $-5.025$ & $3.572$ & $-178.79$ & $-2.913$ & $0.335$ & $-134.17$ & $-4.992$ & $6.784$ & $-603.04$ & $-3.487$\\
      $10$ & $160$ & $0.055$ & $-23.39$ & $-5.025$ & $3.583$ & $-179.32$ & $-2.913$ & $0.336$ & $-134.56$ & $-4.992$ & $6.804$ & $-604.84$ & $-3.487$\\
      $5$ & $320$ & $0.055$ & $-23.30$ & $-5.025$ & $3.572$ & $-178.79$ & $-2.913$ & $0.335$ & $-134.16$ & $-4.992$ & $6.784$ & $-603.04$ & $-3.487$\\
      $2$ & $800$ & $0.053$ & $-22.86$ & $-5.026$ & $3.514$ & $-175.90$ & $-2.913$ & $0.329$ & $-131.99$ & $-4.992$ & $6.674$ & $-593.31$ & $-3.487$\\
      $1$ & $1600$ & $0.051$ & $-22.07$ & $-5.027$ & $3.409$ & $-170.65$ & $-2.913$  & $0.319$ & $-128.05$ & $-4.992$ & $6.476$ & $-575.67$ & $-3.487$\\
      \bottomrule\bottomrule
    \end{tabular}
  }
  \caption{Temperature $T$, chemical potential $\mu$, and entropy per mode number $S/M$ as a function of the number of waveguides along each direction $M_1$, $M_2$ for two-dimensional rectangular waveguide arrays with $\kappa_1=1$, $\kappa_2=4$. The power per mode number and energy per mode number are $N/M=1/400$, $U/M=-3/800$ in columns 3-5, $N/M=1/50$, $U/M=-3/800$, in columns 6-8, $N/M=1/400$, $U/M=-1/1600$ in columns 9-11, and $N/M=1/50$, $U/M=-1/1600$ in columns 12-14.\label{tab:4}}
\end{table*}

\begin{table*}
  \centerline{
    \small
    \begin{tabular}{rr|rrr|rrr|rrr|rrr} \toprule\toprule
      \multicolumn{2}{c}{$\kappa_1=1,\kappa_2=0.8$}\vline & \multicolumn{3}{c}{$\frac NM=\frac{1}{400}$, $\frac UM=-\frac{3}{800}$}\vline  &\multicolumn{3}{c}{$\frac NM=\frac{1}{50}$, $\frac UM=-\frac{3}{800}$}\vline & \multicolumn{3}{c}{$\frac NM=\frac{1}{400}$, $\frac UM=-\frac{1}{1600}$}\vline & \multicolumn{3}{c}{$\frac NM=\frac{1}{50}$, $\frac UM=-\frac{1}{1600}$} \\\midrule
      $M_1$ & $M_2$ & $T$ & $\mu$ & $S/M$ & $T$ & $\mu$ & $S/M$ & $T$ & $\mu$ & $S/M$ & $T$ & $\mu$ & $S/M$\\\midrule
      $400$ & $1$ & $0.001$ & $-2.08$ & $-5.820$ & $0.211$ & $-10.73$ & $-2.921$ & $0.020$ & $-8.11$ & $-5.007$ & $0.404$ & $-35.94$ & $-3.488$ \\
      $200$ & $2$ & $0.004$ & $-3.03$ & $-5.438$ & $0.281$ & $-14.22$ & $-2.919$ & $0.026$ & $-10.77$ & $-5.003$ & $0.533$ & $-47.40$ & $-3.488$ \\
      $100$ & $4$ & $0.005$ & $-3.51$ & $-5.363$ & $0.314$ & $-15.90$ & $-2.918$ & $0.030$ & $-12.05$ & $-5.002$ & $0.595$ & $-52.98$ & $-3.488$ \\
      $50$ & $8$ & $0.005$ & $-3.67$ & $-5.342$ & $0.329$ & $-16.65$ & $-2.918$ & $0.031$ & $-12.62$ & $-5.002$ & $0.624$ & $-55.51$ & $-3.488$ \\
      $40$ & $10$ & $0.005$ & $-3.69$ & $-5.339$ & $0.332$ & $-16.77$ & $-2.918$ & $0.031$ & $-12.71$ & $-5.002$ & $0.628$ & $-55.90$ & $-3.488$ \\
      $20$ & $20$ & $0.006$ & $-3.71$ & $-5.337$ & $0.333$ & $-16.85$ & $-2.918$ & $0.031$ & $-12.77$ & $-5.001$ & $0.631$ & $-56.16$ & $-3.488$ \\
      $10$ & $40$ & $0.005$ & $-3.67$ & $-5.343$ & $0.326$ & $-16.49$ & $-2.918$ & $0.031$ & $-12.50$ & $-5.002$ & $0.617$ & $-54.93$ & $-3.488$ \\
      $5$ & $80$ & $0.005$ & $-3.53$ & $-5.365$ & $0.306$ & $-15.51$ & $-2.918$ & $0.029$ & $-11.77$ & $-5.002$ & $0.580$ & $-51.62$ & $-3.488$ \\
      $4$ & $100$ & $0.005$ & $-3.43$ & $-5.380$ & $0.296$ & $-14.99$ & $-2.918$ & $0.028$ & $-11.38$ & $-5.003$ & $0.561$ & $-49.88$ & $-3.488$ \\
      $2$ & $200$ & $0.003$ & $-2.78$ & $-5.510$ & $0.243$ & $-12.34$ & $-2.920$ & $0.023$ & $-9.38$ & $-5.005$ & $0.460$ & $-40.99$ & $-3.488$ \\
      $1$ & $400$ & $0.000$ & $-1.60$ & $-7.108$ & $0.134$ & $-6.90$ & $-2.926$ & $0.012$ & $-5.23$ & $-5.016$ & $0.258$ & $-23.01$ & $-3.489$ \\
      $1600$ & $1$ & $0.001$ & $-2.08$ & $-5.819$ & $0.211$ & $-10.75$ & $-2.921$ & $0.020$ & $-8.12$ & $-5.007$ & $0.404$ & $-36.01$ & $-3.488$ \\
      $800$ & $2$ & $0.004$ & $-3.04$ & $-5.437$ & $0.281$ & $-14.26$ & $-2.919$ & $0.026$ & $-10.80$ & $-5.003$ & $0.534$ & $-47.53$ & $-3.488$ \\
      $320$ & $5$ & $0.005$ & $-3.59$ & $-5.352$ & $0.323$ & $-16.32$ & $-2.918$ & $0.030$ & $-12.37$ & $-5.002$ & $0.611$ & $-54.39$ & $-3.488$ \\
      $160$ & $10$ & $0.006$ & $-3.70$ & $-5.336$ & $0.336$ & $-16.97$ & $-2.918$ & $0.032$ & $-12.86$ & $-5.001$ & $0.636$ & $-56.58$ & $-3.488$ \\
      $80$ & $20$ & $0.006$ & $-3.74$ & $-5.330$ & $0.341$ & $-17.25$ & $-2.918$ & $0.032$ & $-13.07$ & $-5.001$ & $0.646$ & $-57.51$ & $-3.488$ \\
      $40$ & $40$ & $0.006$ & $-3.75$ & $-5.329$ & $0.342$ & $-17.28$ & $-2.918$ & $0.032$ & $-13.10$ & $-5.001$ & $0.648$ & $-57.63$ & $-3.488$ \\
      $20$ & $80$ & $0.006$ & $-3.73$ & $-5.332$ & $0.338$ & $-17.10$ & $-2.918$ & $0.032$ & $-12.96$ & $-5.001$ & $0.641$ & $-57.02$ & $-3.488$ \\
      $10$ & $160$ & $0.005$ & $-3.69$ & $-5.341$ & $0.329$ & $-16.62$ & $-2.918$ & $0.031$ & $-12.60$ & $-5.002$ & $0.622$ & $-55.37$ & $-3.488$ \\
      $5$ & $320$ & $0.005$ & $-3.54$ & $-5.363$ & $0.308$ & $-15.57$ & $-2.918$ & $0.029$ & $-11.82$ & $-5.002$ & $0.583$ & $-51.84$ & $-3.488$ \\
      $2$ & $800$ & $0.003$ & $-2.78$ & $-5.509$  & $0.244$ & $-12.36$ & $-2.920$ & $0.023$ & $-9.40$ & $-5.005$ & $0.461$ & $-41.07$ & $-3.488$ \\
      $1$ & $1600$ & $0.000$ & $-1.60$ & $-7.104$ & $0.135$ & $-6.92$ & $-2.926$ & $0.012$ & $-5.24$ & $-5.016$ & $0.259$ & $-23.05$ & $-3.489$ \\
      \bottomrule\bottomrule
    \end{tabular}
  }
  \caption{Temperature $T$, chemical potential $\mu$, and entropy per mode number $S/M$ as a function of the number of waveguides along each direction $M_1$, $M_2$ for two-dimensional rectangular waveguide arrays with $\kappa_1=1$, $\kappa_2=0.8$. The power per mode number and energy per mode number are $N/M=1/400$, $U/M=-3/800$ in columns 3-5, $N/M=1/50$, $U/M=-3/800$, in columns 6-8, $N/M=1/400$, $U/M=-1/1600$ in columns 9-11, and $N/M=1/50$, $U/M=-1/1600$ in columns 12-14.\label{tab:5}}
\end{table*}

In the one-dimensional case ($M_2=1$, $M=M_1$), the numerical results shown in Table~\ref{tab:1} confirm our theoretical predictions of Section~\ref{subsec:stress} that such lattices are extensive. Even in the case of very small $M$, such as $M=5$ or even $M=2$ (where thermodynamic theory
does not apply), only very small deviations in the entropy per mode number are observed.

In Table~\ref{tab:2} the coupling coefficients are selected to be the same in both directions $\kappa_1=\kappa_2=1$. The four different cases shown in the columns correspond to the four different combinations of low or high power with low or high energy. In the rows we vary $M_1$ and $M_2$ while keeping their product $M=M_1M_2$ constant and then follow the same procedure for different values of $M$. We start to observe small deviations in $S_M$ as we approach the one-dimensional limit. The low-power and high-energy case exhibits the largest fluctuations. However, as we increase the power or decrease the energy these deviations become minimal even close to the one-dimensional limit. We select the reference value for the entropy from the simulation with the largest number of modes with $M_1=M_2$, or $S_\mathrm{ref}=S(\sqrt{M_\mathrm{max}},\sqrt{M_\mathrm{max}})$. We define the percentage variation in the entropy from the reference value as
\[
  \Var(M_1,M_2)=
  100\frac{|S(M_1,M_2)-S_\mathrm{ref}|}
  {|S_\mathrm{ref}|}.
\]
For example, in Table~\ref{tab:2} the reference value is $S(40,40)$ and,
in the low-power and high-energy case
$\Var(400,4)=0.6\%$,
in the high-power high-energy case
$\Var(400,4)=0.0179\%$,
in the low-power low-energy case
$\Var(400,4)=0.0185\%$,
and finally in the high-power high-energy case
$\Var(400,4)=0.0013\%$.
Deviations from extensivity give rise to stress and strain corrections. 

In Tables~\ref{tab:3} and~\ref{tab:4}, we present results for anisotropic coupling coefficients $\kappa_1\neq\kappa_2$. Specifically, in Table~\ref{tab:3} $\kappa_1=1$ and $\kappa_2=2$, while in Table~\ref{tab:4} $\kappa_1=1$ and $\kappa_2=4$. The selected values of the power and the energy are the same as in Table~\ref{tab:2}. Increasing the coupling coefficient along one-direction promotes coupling, which reflects to increased temperatures. One might expect that increasing the coupling anisotropy is going to result to larger deviations from the extensive character of the entropy. However, we do not observe any noticeable changes in $S/M$ as we vary $M_1$ and $M_2$ in comparison to Table~\ref{tab:2}.

In Table~\ref{tab:5} we select $\kappa_1=1$ and $\kappa_2=0.8$ while the rest of the parameters remain the same. We observe more noticeable changes in $S_M$, especially close the 1D limit. However, these coupling coefficients lead to temperatures very close to zero (condensation limit) especially in the low power and high energy case and require special treatment.

\section{Conservation laws of the discrete nonlinear Schr\"odinger equation with $z$-dependent coupling coefficients\label{sec:app:conserv}}
We consider the discrete nonlinear Schr\"odinger equation with two transverse directions, $z$-dependent coupling coefficients, and cubic nonlinearity 
\begin{equation}
  i\dot{u}_m
  +
  \kappa_1(z)\Delta_1u_m
  +
  \kappa_2(z)\Delta_2u_m
  + 
  \gamma |u_{m}|^2u_{m}=0,
  \label{eq:dnls}
\end{equation}
where $m=(m_1,m_2)$, $m_1=1,\ldots,M_1$, and $m_2=1,\ldots,M_2$. Multiplying with $u_{m}^*$ we obtain 
\begin{equation*}
  iu_m^*\dot{u}_m
  +
  \kappa_1(z)u_m^*\Delta_1u_m
  +
  \kappa_2(z)u_m^*\Delta_2u_m
  + 
  \gamma |u_{m}|^4=0.
\end{equation*}
Complex conjugating the above equation, assuming that the coupling coefficients are real, and subtracting the two expressions we obtain
\begin{multline*}
  i\frac{\dif}{\dif z}|u_m|^2
  +
  \kappa_1(z)(u_m^*\Delta_1u_m-u_m\Delta_1u_m^*)
  +\\
  \kappa_2(z)(u_m^*\Delta_2u_m-u_m\Delta_2u_m^*)
  =0.
\end{multline*}
Taking the sum over $m$ we get 
\begin{multline*}
  i\frac{\dif}{\dif z}N
  +
  \sum_m\kappa_1(z)(u_m^*\Delta_1u_m-u_m\Delta_1u_m^*)
  +\\
  \sum_m\kappa_2(z)(u_m^*\Delta_2u_m-u_m\Delta_2u_m^*)
  =0,
\end{multline*}
where
\[
  N = \sum_m|u_m|^2
\]
is the total power. For zero boundary conditions $u_{M_1+1,m_2}=u_{0,m_2}=u_{m_1,M_2+1}=u_{m_1,0}=0$ we have that
\[
  \sum_m u_m^*\Delta_ju_m=\sum_mu_m\Delta_ju_m^*,\quad
  j=1,2
\]
and thus $\dot N = 0$ meaning that $N$ remains constant.

To find how the Hamiltonian varies when the coupling coefficients are $z$-dependent, we multiply Eq.~(\ref{eq:dnls}) with $\dot u_m^*$ and thus
\begin{multline*}
  i|\dot{u}_m|^2
  +
  \dot u_m^*\kappa_1(z)\Delta_1u_m
  +
  \dot u_m^*\kappa_2(z)\Delta_2u_m
  + \\
  \gamma |u_{m}|^2u_{m}\dot u_m^*=0,
\end{multline*}
Complex conjugating the above equation and adding the two equations we obtain
\begin{multline*}
  \dot u_m^*\kappa_1(z)\Delta_1u_m
  +
  \dot u_m\kappa_1(z)\Delta_1u_m^*
  + \\
  \dot u_m^*\kappa_2(z)\Delta_2u_m
  +
  \dot u_m\kappa_2(z)\Delta_2u_m^*
  + 
  \frac\gamma2\frac{\dif |u_{m}|^4}{\dif z} =0.
\end{multline*}
Summing over $m$ leads to
\begin{multline*}
    \frac{\dif}{\dif z}
  \sum_m 
  \left[
    \kappa_1u_m^*\Delta_1 u_m
    +
    \kappa_2u_m^*\Delta_2 u_m
    +
    \frac\gamma2|u_m|^4
  \right]
  = \\
  \sum_m\left[
    \dot\kappa_1\Delta_1 u_mu_m^*
    +
    \dot\kappa_2\Delta_2 u_mu_m^*
  \right].
\end{multline*}
Defining by
\[
  R_j = \sum_mu_m^*\Delta_ju_m,\quad j=1,2,
\]
\[
  H_{NL} =\frac\gamma2\sum_m|u_m|^4,
\]
and
\begin{equation}
  H = R_1\kappa_1+R_2\kappa_2+H_{NL},
  \label{eq:appD:H}
\end{equation}
we derive
\[
  \frac{\dif H}{\dif z} =
  \frac{\dif\kappa_1}{\dif z}H_1+
  \frac{\dif\kappa_2}{\dif z}H_2.
\]
Multiplying with $\dif z$ we find that the variations to the value of the Hamiltonian are proportional to the variations in the coupling coefficients 
\begin{equation}
  \dif H = R_1\dif\kappa_1+R_2\dif\kappa_2. 
  \label{eq:appD:difH}
\end{equation}
By taking the differential of Eq.~(\ref{eq:appD:H}) and subtracting Eq.~(\ref{eq:appD:difH}) we derive
\begin{equation}
  \kappa_1\dif R_1+\kappa_2\dif R_2=0.
\end{equation}

\newcommand{\noopsort[1]}{} \newcommand{\singleletter}[1]{#1}

\end{document}